# Ultrafast Broadband Photodetectors based on Three-dimensional Dirac Semimetal $Cd_3As_2$

Qinsheng Wang[1,#], Cai-Zhen Li[2,#], Shaofeng Ge[1], Jin-Guang Li[2], Wei Lu[1], Jiawei Lai[1], Xuefeng Liu[1], Junchao Ma[1], Da-Peng Yu[2,3], Zhi-Min Liao[2,3,*], Dong Sun[1,3,*]

[1]International Center for Quantum Materials, School of Physics, Peking University, Beijing 100871, P. R. China
[2]State Key Laboratory for Mesoscopic Physics, School of Physics, Peking University, Beijing 100871, P.R. China
[3]Collaborative Innovation Center of Quantum Matter, Beijing 100871,P. R. China
[#] These authors contribute equally to this work.
[*]Addess correspondence to: sundong@pku.edu.cn (D.S.); liaozm@pku.edu.cn (Z.M.L.)

**Abstract:**
**The efforts to pursue photo detection with extreme performance in terms of ultrafast response time, broad detection wavelength range, and high sensitivity have never been exhausted as driven by its wide range of optoelectronic and photonic applications such as optical communications, interconnects, imaging and remote sensing[1]. 2D Dirac semimetal graphene has shown excellent potential toward high performance photodetector with high operation speed, broadband response and efficient carrier multiplications benefiting from its linear dispersion band structure with high carrier mobility and zero bandgap[2-4]. As the three dimensional analogues of graphene, Dirac semimetal $Cd_3As_2$ processes all advantages of graphene as a photosensitive material but potentially has stronger interaction with light as bulk material and thus enhanced responsivity[5,6], which promises great potential in improving the performance of photodetector in various aspects . In this work, we report the realization of an ultrafast broadband photodetector based on $Cd_3As_2$. The prototype metal-$Cd_3As_2$-metal photodetector exhibits a responsivity of 5.9 mA/W with response time of about 6.9 ps without any special device optimization. Broadband responses from 0.8 eV to 2.34 eV are measured with potential detection range extendable to far infrared and terahertz. Systematical studies indicate that the photo-thermoelectric effect plays important roles in photocurrent generation, similar to that in graphene. Our results suggest this emerging class of exotic quantum materials can be harnessed for photo detection with high sensitivity and high speed (~145 GHz) in challenging middle/far-infrared and THz range.**

Three dimensional Dirac semimetal $Cd_3As_2$[7,8] is a stable compound with ultrahigh carrier mobility up to $9\times10^6$ $cm^2V^{-1}S^{-1}$ (refs. 9-11), as a result of supressed backscattering of high Fermi velocity 3D Dirac fermions[9]. The high mobility of $Cd_3As_2$ surpasses suspended graphene and any bulk semiconductors, promising for new electronics and optoelectronics with supereme performance[5, 6]. Comparing to their two dimensional counterpart graphene[5,6,12,13] and surface state of topological insulator[14], 3D Dirac semimetals are more robust against enviromental defects or excess conducting bulk electrons[11]. On the other hand, 3D Dirac semimetals possess all advantages of 2D Dirac semimetals as photosensitive materials, which is inherent from the gapless linear dispersion of massless Dirac Fermions: extermely high mobility[9] and ultrafast transient time[15] for high speed response approaching terahertz operation speed[4,16]; gapless bandstructure for chanlleging low energy photon detection[17] down to THz frequency[18-20] and efficient carrier multiplications to enhance the internal quantum efficiency[21,22]. Consequently, the emergency of stable 3D Dirac semimetal $Cd_3As_2$ provides outstanding opportunity as new class of material platform for optoelectronics.

Experimental studies on $Cd_3As_2$ so far mainly focus on the tranport and angle resolved photoemission spectroscopy (ARPES) measurements to confirm 3D Dirac semimetal phase and its related electronic behavior near the Fermi level, such as giant magnetoresistance (MR), non-trivial quantum oscillations and landau level splitting under magnetic field[10,23-25]. Despite the potential exceptional optical properties, the photonic and optoelectronical response of 3D Dirac semimetal is largely unexplored[26]. Here, we investigate the use of $Cd_3As_2$ as photodetectors ultilizing metal-$Cd_3As_2$-metal structure (device schematic shown in Fig. 1a, b) and characterize the performance of the photodetectors with spatially resolved scanning photocurrent microscopy (SPCM) and time resolved photocurrent spectroscopy (TRPC). On light absorption, the photo-excited carriers in $Cd_3As_2$ would normally relax on a timescale of tens of picoseconds[27,28]. As the photocurrent (PC) response is ultrafast (~6.9 ps) as revealled by time resolved PC measurement in this work, hot carriers dominated photothermoelectric effect play crutial role on PC geneation in $Cd_3As_2$ similar to that in graphene[4,29]. Combining with its broad spectrum response and reasonable responsivity, this photodetectors may find a wide range of photonic applications including high-speed optical communications, interconnects, terahertz detection, imaging, remote sensing, surveillance and spectroscopy[30,31].

Figure 1 shows the schematic of PC measurement and sample characterization. The $Cd_3As_2$ nanowires (Fig. 1c) and nanoplates (Fig. 1g) were synthesized by chemical vapor deposition (CVD) method[25,32]. The lateral metal-$Cd_3As_2$-metal devices based on individual $Cd_3As_2$ nanowires (Fig. 1d) and nanoplates (Fig. 1h) were fabricated for photodetection. The diameters of the nanowires range from 100 to 200 nm and the channel length of a typical device shown in Fig. 1d is about 7~8 μm. The linear current-voltage curve shown in Fig. 1e indicates the Ohmic contacts between the nanowire and electrodes, and the resistance of the device ~ 5.37 kΩ is extracted. The nanoplates are typically 200 nm thick with a length of 5~15 μm and a width of 3~4 μm. A typical nanoplate device shown in Fig. 1h has source-drain resistance of 606 Ω (Fig. 1i), an order smaller than that of nanowire device. Both the nanoplates and nanowires are grown along the (112) direction according to the selected area electron diffraction (Supplementary Fig. S1). Figures 1f, 1j show the room temperature SPCM images of a nanowire and nanoplate device with relatively low excitation power, respectively. While the photo response of the nanowire device is along the whole nanowire, the photocurrent of the nanoplate devices is mainly at the interface of the $Cd_3As_2$-metal junctions.

To study the PC generation mechanism on $Cd_3As_2$ photodetectors, excitation power dependent measurements were performed on nanowire (Fig. 2a-g) and nanoplate devices (Supplementary Fig. S4). For nanowire devices, the power dependence of PC responses show different patterns along the nanowire under two different excitation power ranges: under low excitation energy (< 60 μW), there is a global positive photocurrent response along the entire nanowire (Fig. 2 c,d); under excitation power above 60 μW, the response becomes bipolar (Fig. e, f). Within each excitation range, the PC response generally increases with excitation power (Fig. g). However, for nanoplate device, the SPCM pattern stays qualitatively the same with excitation power increased up to 800 μW (Supplementary Fig. S4).

The stark contrast in nanowire and nanoplate devices implies the photo-thermoelectric (PTE) effect may dominate over photovoltaic (PV) effect in PC response of $Cd_3As_2$, as the PV effect alone cannot interpret the power and temperature dependent SPCM responses of devices. Photocurrent generation from PV effect is based on separation of photo generated electron-hole (e-h) pairs by built-in electric fields at junctions between differently doped sections. In a metal-$Cd_3As_2$-metal photodetector used in this work, built-in electric field can be formed at the $Cd_3As_2$-metal junction due to the work-function difference between $Cd_3As_2$ and the gold electrode. The PV generated photocurrent should reverse sign between $Cd_3As_2$-metal and metal-$Cd_3As_2$ junctions as observed in a nanoplate device, but this is in contradiction to that is observed in a nanowire devices: globally positive over the whole nanowire under low excitation power. This contradiction implies effects other than PV have to be taken into account for polarity reversal of PC response in $Cd_3As_2$ nanowire device under weak excitations.

In a nanoplate device, at room temperature, the photocurrent generation is mainly around the two overlapped corner of $Cd_3As_2$ and metal contact as shown in SPCM (Fig. 2h,i,j). The SPCM measurements of nanoplate devices with different geometric shapes indicate the PC response pattern strongly depends on the geometry of $Cd_3As_2$ sample and metal contact (Supplementary Fig. S3). Furthermore, in temperature dependent SPCM measurement of a nanoplate device (Fig. 2k,l), as the temperature cools down from room temperature down to 15 K, the PC response pattern changes prominently: at low temperature, the negative photocurrent dominates the PC pattern; as the temperature increases, the peak position of negative response move towards the contact and the areas of positive/negative parts gradually become equal. The above sample geometry and temperature dependent results cannot be interpreted with sole PV effect either, as the PC generation should be always along the interface between $Cd_3As_2$ and metal contact where the built-in electric field exists, in a pure PV effect picture.

As PV effect alone cannot interpret the experimentally observed results, PTE effect may come into play. For PTE effect, the photo generated hot carriers can produce a photo voltage through the electron temperature gradient and thermoelectric power (Seebeck coefficient S) difference of two adjacent regions. Generally, $V_{PTE}$ can be calculated by integrating the optically induced electron temperature gradient together with a spatially varying Seebeck coefficient: $V_{PTE}=\int S \cdot \nabla T_e dx$, where S is carrier density dependent function. As the $SiO_2$ substrate does not conducting heat as efficient as metal and semimetallic $Cd_3As_2$, the gold contact is the major heat dissipation channel of the device. So, after photon excitation, it's the steady state electron temperature distribution over the $Cd_3As_2$ sample that determines the PC generated from PTE effect. While quantitatively understanding the dependence of the observed SPCM pattern on $Cd_3As_2$ photodetectors requires the knowledge of the specific heats of electrons and phonons and thermoelectric coefficient at elevated

$T_e$, which is not available at present. Qualitatively, the electron temperature distribution over $Cd_3As_2$ sample, and thus the PTE generated PC, should strongly depend on excitation positions (distance from the two metal electrodes and edge of $Cd_3As_2$), detailed sample shape, excitation power and initial adjacent electron/lattice temperature around the excitation position. As the nanowire device is less than 200 nm in diameter, the heat transport is quasi one-dimensional along the nanowire, and excitation power may easily reach the threshold to change the polarity of the temperature gradient that is created on the sample, thus the SPCM pattern along the device can be modified significantly by increasing excitation power as observed in Fig. 2g. For nanoplate device with width of 3~4 μm, the heat dissipation is two dimensional, which is more efficient comparing to that in a nanowire, so the SPCM pattern stays the same with excitation power over 800 μW (Supplementary Fig. S4). However, for temperature dependent measurement, the initial adjacent electron/lattice temperature around the excitation position can affect the SPCM pattern as observed in Fig 2 k,l.

The dominating role of PTE effect in photo detection of $Cd_3As_2$ is similar to graphene[4,29,33,34]. In graphene, photo-excited hot carriers remain at a temperature higher than lattice for several picoseconds (transient time of hot carrier). As this timescale is comparable to the photo response time of graphene photodetector, hot-carrier-assisted transport plays an important role in photo response of graphene and contributes to PTE effect significantly. As the transient time of photo-excited hot carriers is measured to be on the order of picoseconds in $Cd_3As_2$[28], the efficient PTE effect implies the photo response time of $Cd_3As_2$ photodetector can also be very fast. Figure 3 shows the TRPC measurement result of $Cd_3As_2$ nanoplate device at room temperature with cross-polarized pump-probe configuration (co-polarized configuration shown in Figure S2). The pump-probe pulse width can be deducted from the autocorrelation that is measured from reflection measurement, which shows an oscillation with a FWHM of 0.35 ps, indicating the autocorrelation of pulse width of the laser is on the order of 200 fs. The exponential decay time constant of the pump induced photocurrent modification dip is about 6.87 ps, which is equivalent to an intrinsic bandwidth of 145 GHz. This speed is comparable to graphene and outperform other technologies being investigated for optical communications, such as monolithically integrated Ge[35,36].

To further elaborate the advantage of gapless linear dispersion bandstructure of $Cd_2As_3$ in photo detection, the broadband wavelength responses of the device are measured. Figure 4a shows the short-circuit photocurrent responses under different excitation photon energy from 0.8 eV to 2.33eV by switching on/off the laser illumination of 10-μW pulse (6 ps, 20 MHz), taking from positive maximum PC position of the device at each wavelength. We note the photon energy range is limited by the available light source rather than response of the device itself. As a zero gap Dirac semimetal with ultrahigh mobility, $Cd_3As_2$ should possess ultra-broadband response extendable to much lower photon energy[37] down to THz similar to that of graphene[18-20]. Wavelength dependent SPCM measurements (Fig. 4b for 532 nm and Fig. 4c for 800 nm) indicate the PC response patterns are similar with different excitation photon energies at low power excitation. Regarding to the responsivity, it is about 2.0 mA/W under pulse excitation of 633 nm as shown in Fig 4d. With CW excitation at 633 nm on a different nanoplate device, higher responsivity of 5.88 mA/W can be achieved. This responsivity is about 10 times larger than that of a graphene detector (0.5 mA/W) with a similar device structure and experimental conditions[2]. Although much higher responsivity has already been achieved in graphene with special device design and optimization[3,6], the 3D $Cd_3As_2$ certainly outperforms in terms of responsivity due to stronger light interaction comparing to one atomic thick graphene.

Furthermore, the responsivity of the detector exhibits strong excitation photon energy dependence (Fig. 4d): for example, the responsivity at 1.5 eV is about 10 times larger than that at 0.8 eV. Although many effects, such as wavelength dependent focusing spot size, spatial non-uniformity of the response and wavelength dependent light absorption coefficient, may contribute to the excitation photon energy dependence of the observed responsivities, the overall rapid responsivity enhancement with photon energy may imply efficient carrier/hot carrier multiplication process in $Cd_3As_2$ considering its gapless band structure and efficient electron-electron scattering. Similar to that in graphene, electron-electron scattering can lead to the conversion of one high-energy e-h pair into multiple e-h pairs of lower energy. This process, also denoted as carrier multiplication can potentially enhance the quantum efficiency of photo detection. In addition to carrier multiplication, the hot carrier multiplication process can also be very efficient in contribution to hot carrier assisted PTE response in $Cd_3As_2$[38]. These results together promise highly efficient broadband extraction of light energy into electronic degrees of freedom, enabling high-efficiency optoelectronic applications.

In summary, we have demonstrated high performance photodetectors based on 3D Dirac semimetal $Cd_3As_2$. The unique properties of $Cd_3As_2$ enable high bandwidth (~145 GHz), broad wavelength range, zero biased and PTE dominated photodetection. With rapid development of large scale thin film growth by molecular beam epitaxial (MBE) and large scale integration possibility[26], 3D Dirac semimetal materials promise enormous potential for ultrafast and ultrasensitive detection of light in very broad wavelength range including the technically challenging middle/far IR and THz wavelength range, with high quantum efficiency and convenient integration with flexible device platform.

**Methods:**

**Sample growth and device fabrication.** $Cd_3As_2$ nanoplates and nanowires have been synthesized by chemical vapor deposition (CVD) method in a horizontal tube following the same recipe as described in the references[25,32]. The synthesized nanoplates and nanowires were transferred onto a Si substrate with 285 nm $SiO_2$ layer. Using the e-beam lithography and evaporation technique, Au electrodes with 500 nm (200 nm) were deposited on individual $Cd_3As_2$ nanoplates (nanowires) to form metal-$Cd_3As_2$-metal photodetectors.

**Sample characterization.** The SEM characterizations were performed in a FEI NanoSEM 430 system. The TEM and EDS characterizations were performed in a FEI Tecnai F20 TEM equipment with energy-dispersive X-ray spectroscopy system.

**Optoelectronic measurements.** Standard scanning photocurrent measurements[33] were performed in ambient conditions using 532nm/800 nm CW laser with ~1 μm spatial resolution. The reflection signal and photocurrent were recorded simultaneously to get the reflection and photocurrent mapping. The laser beam was modulated with a mechanical chopper (379Hz), and the short-circuit photocurrent signal was detected with a current pre-amplifier and a lock-in amplifier. For the temperature dependent measurements, a scanning mirror was used to allow the beam to be scanned on the sample, otherwise, the sample was mounted on a three dimensional piezo stage to be scanned.

For wavelength dependent measurement, a series of bandpass filters were used to select the desired wavelength from white light supercontinuum output of Fianium: WhiteLase-Micro (20 MHz, 6 ps, 450 nm-2200 nm). The laser was focused on the same place of the device by a 100X NIR objective.

Time resolved photocurrent measurement was carried out with 800-nm pulses generated by a Coherent MIRA laser with repetition rate of 76 MHz. The output of Mira was split into two optical paths, one used as the probe (chopped) to record the short-circuit photocurrent response, and the other used as pump containing a linear delay stage to vary the pump-probe delay. The pulse width is about 200 fs at the sample. A 5X objective was used to focus the laser beam to a spot diameter of about 20 μm which covers the whole sample to minimize the possible effect of sample drift during the delay time scan.

**Extraction of the photo response time.** The fitting of the time-resolved photocurrent (PC) signals was performed using the equation

$$\frac{PC(\Delta t)}{PC(\Delta t \to \infty)} = 1 - A \exp(-\frac{|\Delta t|}{\tau})$$

where amplitudes A and time constants τ are the fitting parameters. The exponential time constant τ give the intrinsic response time of $Cd_3As_2$ photodetector.

**Acknowledgement:** This project has been supported by the National Basic Research Program of China (973 Grant Nos. 2016YFA0300802, 2014CB920900 and 2012CB921300), the National Natural Science Foundation of China (NSFC Grant Nos. 11274015), the Recruitment Program of Global Experts, Beijing Natural Science Foundation (Grant No. 4142024) and the Specialized

Research Fund for the Doctoral Program of Higher Education of China (Grant No.20120001110066).

**Author contributions**

D.S. conceived and designed the experiments. C.Z.L. and J.G.L. performed the sample growth and device fabrication supervised by Z.M.L. and D.P.Y.. The optoelectronic measurements were performed by Q.W., S. Ge, J. L., W. L., X.L. and J.M., supervised by D.S.. D.S., Z.M.L., Q.W., C.Z.L. and J.W.L. performed the data analysis. All authors discussed the results and contributed to writing the manuscript.

**Additional information**

The authors declare no competing financial interests. Supplementary information accompanies this paper at www.nature.com/naturenanotechnology. Reprints and permission information is available online at http://www.nature.com/reprints. Correspondence and requests for materials should be addressed to D. S. and Z.M.L.

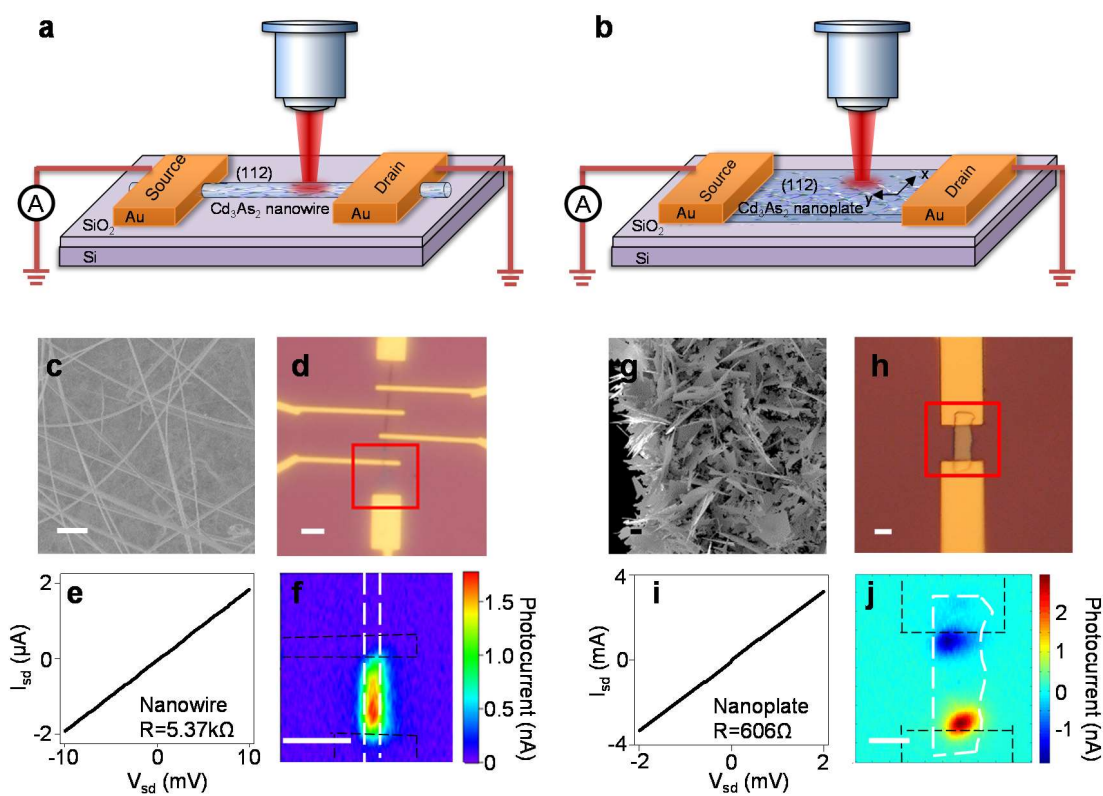

**Figure 1 | Schematic of photocurrent measurement and Cd$_3$As$_2$ sample characterization. a,** Schematic of scanning photocurrent measurement of Cd$_3$As$_2$ nanowire and **b**, nanoplate devices. **c**, SEM image of Cd$_3$As$_2$ nanowires. **d**, Optical image of a typical Cd$_3$As$_2$ nanowire device. **e**, Measured I-V curve and **f,** photocurrent response of a Cd$_3$As$_2$ nanowire device. The photo-excitation power is 10μW at a wavelength of 532nm. **g**, SEM image of Cd$_3$As$_2$ nanoplates. **h**, Optical image of a typical Cd$_3$As$_2$ nanoplate device. **i**, Measured I-V curve and photocurrent response of a Cd$_3$As$_2$ nanoplate device. The photo-excitation power is 25μW at 800nm. Both the I-V curves and the photocurrent responses are measured at room temperature. The red solid rectangular in **d** and **h** indicate the area where the photocurrent are measured, and the black and white dashed lines in **f** and **j** indicate the outline of electrode and sample, respectively.

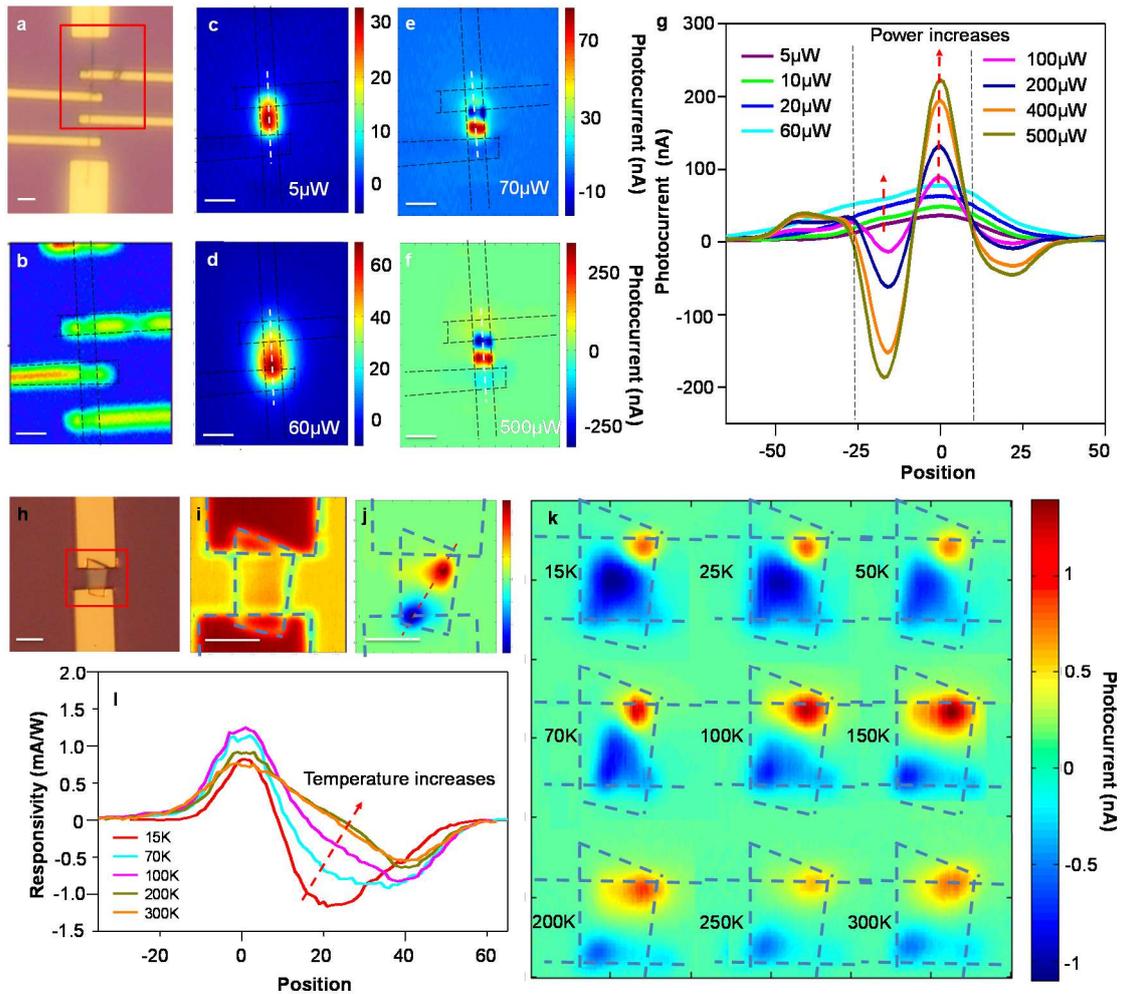

**Figure 2 | Photo-thermoelectric outputs in Cd₃As₂ based photodetectors. a (h),** Optical micrograph of Cd₃As₂ nanowire (nanoplate) devices, where the red box indicates the scanning area of scanning photocurrent microscopy (SPM) measurement. **b (i),** Scanning reflection micrograph of the Cd₃As₂ nanowire (nanoplate) device. Dashed line indicates the edge of electrodes and Cd₃As₂. **c, d, e, f,** SPM image the Cd₃As₂ nanowire device with excitation power of 5, 60, 70, 500 μW at 532 nm, respectively. Dashed line marks the profile of the device. **g,** Line cut of photocurrent response along centerline of nanowire under different excitation power. Dashed line marks the edge of contact. **k,** SPM of the Cd₃As₂ nanoplate device with excitation power of 50 μW at 800 nm at the temperature of 15, 25, 50, 70, 100, 150, 200, 250, 300K, respectively. Dashed line marks the profile of the device. **l,** Line cut of photocurrent response of nanoplate under different temperatures along solid red line in the figure **j**. The scale bar is 5 μm in all figures.

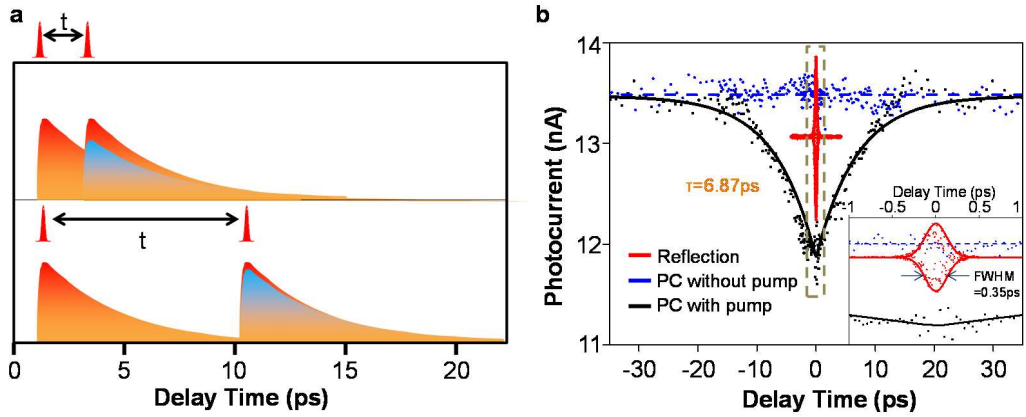

**Figure 3 | Time-resolved photocurrent measurement of $Cd_3As_2$ photodetectors. a,** Schematic of pump-probe photocurrent measurement. The probe beam is mechanically chopped and thus only the probe induced photocurrent is recorded by a lock-in amplifier. Here we compare two scenarios: independent photocurrent responses induced by individual pump or probe (orange area) and reduced photocurrent response of probe (blue area) due to the presence of the pump at different delay times. **b,** Probe induced photocurrent with pump on (black) and off (blue) as a function of pump–probe pulse delay. Both pump and probe power are 8.7 mW with the spot diameter of 20 μm. The pump and probe beams are cross polarized to minimize the interference at timezero. The smooth solid lines are fitting curve with time constant of 6.87 ps with function mentioned in methods. The red curve is an auto-correlation measurements of the pulse with pulse width of 200 fs in co-polarized configuration. The inset is zoom-in view of the dash box with a FWHM of 0.35 ps of the envelope.

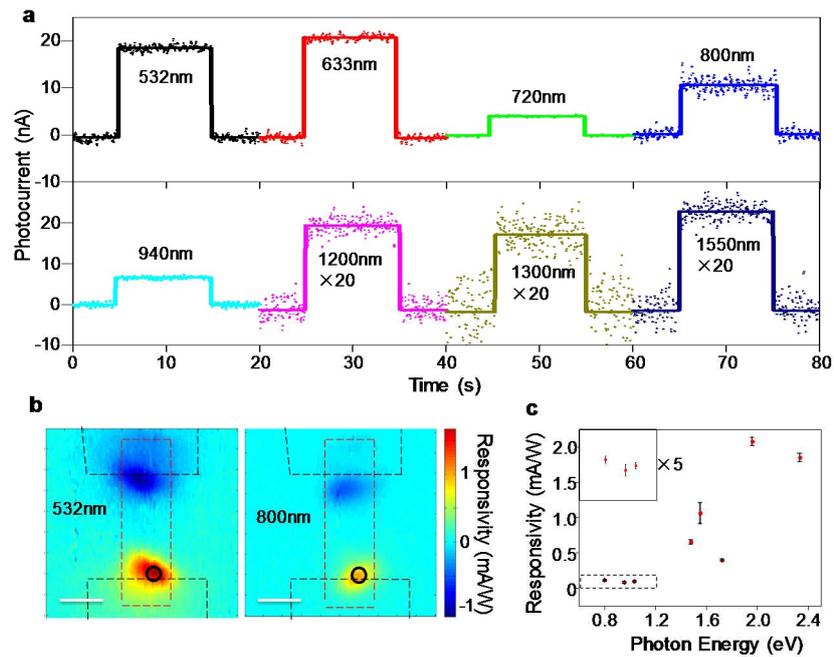

**Figure 4 | Broadband photo response of Cd$_3$As$_2$ photodetectors. a,** Photo-response of a Cd$_3$As$_2$ nanoplate device under different photoexcitation wavelengths (as marked) for the same excitation spot at the device. The excitation power is fixed at 10 μW for all wavelengths. **b,** SPM of the device under 532nm (left) and 800nm (right) excitations. The dashed lines mark the profile of the device. **c,** Photocurrent responsivity as a function of excitation energy of a Cd$_3$As$_2$ nanoplate device.